\documentclass[a4paper,12pt]{article}
\usepackage[pctex32]{graphics}

\begin{document}
\topmargin 0pt
\oddsidemargin 0mm

\newcommand{\alp}{\alpha}
\newcommand{\bta}{\beta}
\newcommand{\gmm}{\gamma}
\newcommand{\del}{\delta}
\newcommand{\omg}{\omega}
\newcommand{\sgm}{\sigma}
\newcommand{\lmd}{\lambda}
\newcommand{\tha}{\theta}
\newcommand{\vph}{\varphi}
\newcommand{\Alp}{\Alpha}
\newcommand{\Bta}{\Beta}
\newcommand{\Gmm}{\Gamma}
\newcommand{\Del}{\Delta}
\newcommand{\Omg}{\Omega}
\newcommand{\Sgm}{\Sigma}
\newcommand{\Lmd}{\Lambda}
\newcommand{\Tha}{\Theta}
\newcommand{\half}{\frac{1}{2}}
\newcommand{\rnd}{\partial}
\newcommand{\nab}{\nabla}

\newcommand{\beqn}{\begin{eqnarray}}
\newcommand{\eeqn}{\end{eqnarray}}
\newcommand{\be}{\begin{equation}}
\newcommand{\ee}{\end{equation}}

\begin{titlepage}

\vspace{5mm}
\begin{center}
{\Large \bf Entropic force and  entanglement system } \vspace{12mm}

{\large   Yun Soo Myung \footnote{e-mail
 address: ysmyung@inje.ac.kr} and Yong-Wan Kim\footnote{e-mail
 address: ywkim65@gmail.com}}
 \\
\vspace{10mm} {\em  Institute of Basic Science and School of
Computer Aided Science, Inje University, Gimhae 621-749, Republic of
Korea}

\end{center}

\vspace{5mm} \centerline{{\bf{Abstract}}}
 \vspace{5mm}
We introduce the isothermal cavity, static holographic screen, and
accelerating surface as holographic screen  to study the entropic
force in the presence of the Schwarzschild black hole. These may
merge to provide a consistent holographic  screen to define the
entropic force on the stretched horizon near the event horizon.
Considering the similarity between the stretched horizon of black
hole and the entanglement system, we may define the entropic force
in the entanglement system without referring the source  mass.

\end{titlepage}
\newpage
\renewcommand{\thefootnote}{\arabic{footnote}}
\setcounter{footnote}{0} \setcounter{page}{2}

\section{Introduction}
Since the discovery of the laws of black hole
thermodynamics~\cite{BCH}, Bekenstein~\cite{Bek} and
Hawking~\cite{Hawk} have suggested a deep connection between
gravity and thermodynamics, realizing black hole entropy and
Hawking radiation.  Later on, Jacobson~\cite{Jac} has demonstrated
that Einstein equations (describing relativistic gravitation)
could be derived by combining general thermodynamic pictures with
the equivalence principle. Padmanabhan~\cite{Pad3} has observed
that the equipartition law for  horizon degrees of freedom
combined with the Smarr formula leads to the Newton's law of
gravity.

Recently, Verlinde has proposed the Newtonian force law as an
entropic force (non-relativistic version)  by using  the holographic
principle and  the equipartition rule~\cite{Ver}.
 If it
is proven correct, gravity is not a fundamental interaction, but
an emergent phenomenon which arises from the statistical behavior
of microscopic degrees of freedom encoded on a holographic screen.
In other words, the force of gravity is not something ingrained in
matter itself, but it is an extra physical effect, emerging from
the interplay of mass, time and space (information).

After his work,  taking the apparent horizon as a holographic
screen to derive  the Friedmann equations~\cite{SG}, derivation of
the Friedmann equations using the equipartition rule and unproved
Unruh temperature~\cite{Pad1,CCO}, the connection to the loop
quantum gravity~\cite{Smo}, the accelerating
surfaces~\cite{makela1}, holographic actions for black hole
entropy~\cite{CM}, and application to holographic dark
energy~\cite{LWh} were considered from the view of the entropic
force.  Furthermore,  cosmological implications were reported in
\cite{Gao,ZGZ,Wang,WLW,LW,LKL}, an extension to the Coulomb
force~\cite{WangT}, and the symmetry aspect of the entropy
force~\cite{Zhao} were investigated. The entropic force was
discussed in the presence of  black
hole~\cite{Myungef,LWW,TW,CCO2}.

However, one of urgent issues to resolve is to answer a question
of {\it how one can construct a spherical holographic screen of
radius $R$ which encloses a source mass $M$ located at the origin
to understand the entropic force.} This is crucial because the
holographic screen (an exotic description of spacetime) originates
from relativistic approaches to black hole ~\cite{Hoo,Suss} and
cosmology~\cite{Bou}. Verlinde has introduced this screen  by
analogy with an absorbing process of a particle around the event
horizon of black hole.  Considering a smaller test mass $m$
located at $\Delta x$ away from the screen and getting the change
of entropy on the  screen,  its behavior should resemble that of a
particle approaching a stretched horizon of a black hole, as
described by Bekenstein~\cite{Bek}.

The next question is why the equipartition rule could be applied
to this non-relativistic holographic screen to derive the
Newtonian force law without any justifications.  For black holes,
the equipartition rule becomes the Smarr formula of $E=NT/2=2ST$
when using $N=4S=A/G$ in the natural units of $\hbar=c=k_B=1$ and
$G=l^2_{pl}$. Also it can be derived from the first law of
thermodynamics $dE=TdS$ for the Schwarzschild black hole where the
Komar charge is just the ADM mass $M$.  Most of cosmological
implications  have used the holographic principle (screen) and
equipartition rule to derive the  Friedmann equations. However,
these implications did not explain  clearly how their approach is
related to the appearance of the entropic force because these
belong to the relativistic approach.  Even though the
equipartition rule is available for the Newtonian dynamics, the
holographic principle of $N=A/G$ is not guaranteed to apply to any
non-relativistic situations. In this sense, this issue is closely
related to the first one.

If the above two questions are answered properly, one will make a
further step to understand the entropic force. However, there is
still a gap between non-relativistic and relativistic approaches.
It seems that Verlinde was recycling some ideas for obtaining
Einstein equations due to Jacobson's derivation of Einstein
equations. Also, we would like to mention that he was using
circular reasoning in his equations, by starting out with gravity.

At this time, it seems  hard to discuss an entropic force without
referring a holographic screen.  Hence, we propose that
introducing a holographic screen is a first step to understand the
entropic force.

In this work, we show  how a holographic screen is constructed to
define the entropic force  well by implementing the Schwarzschild
spacetimes.   Here, we introduce four candidates of  the
isothermal cavity, static holographic screen, accelerating
surface, and stretched horizon to study  the entropic force.
Finally, we consider the entanglement system in the flat spacetime
as a promising candidate for a holographic screen to define the
entropic force without referring the mass $M$.   We must say that
our approach is not a complete scheme to understand the entropic
force  because  we have already introduced a given configuration
of gravitational field (relativistic situation), the Schwarzschild
spacetimes to define a holographic screen.

We show that to define the entropic force, three of the isothermal
cavity, static holographic screen, and accelerating surface  might
merge to provide a unified picture for a holographic screen  on
the stretched horizon near the event horizon.  Considering a close
relationship between black hole thermodynamics on the stretched
horizon and the entanglement system, we propose that an entropic
force may be defined  in the entanglement system  without
referring the source mass $M$.

\section{Entropic force}

In this section, we briefly review how the Newtonian  force law
emerges from  entropic considerations.  Explicitly, when a test
particle with mass $m$ is located near a holographic screen with
distance $\Delta x$, the change of entropy on a holographic screen
may take the form \be \label{eq1} \Delta S= 2\pi m \Delta x. \ee
Considering that the entropy of a system depends on the distance
$\Delta x$, an entropic force $F$ could be arisen from the
thermodynamical conjugate of the distance as \be \label{eq2} F
\Delta x=T \Delta S \ee which may be regarded as an indication
that the first law of thermodynamics is realized  on the
holographic screen.  Plugging (\ref{eq1}) into (\ref{eq2}) leads
to an important connection between the entropic force and
temperature \be \label{eq3} F=2\pi m T. \ee In this work,  we use
this connection mainly  to derive the entropic force, after
setting the temperature $T$ on the holographic screen.

Let us assume that the energy $E$ is distributed on a spherical
screen with radius $R$ and  the source mass $M$ is located at the
origin of coordinates. Then, we assume that the equipartition
rule~\cite{Pad2,Pad3}, the equality of  energy and mass, and the
holographic principle, respectively, hold  as \be \label{eq4}
E=\frac{1}{2 }N T,~~E=M,~~N=\frac{A}{G}=4S \ee with the area of a
holographic screen $A=4\pi R^2$.  These are combined to provide
the temperature on the screen \be \label{eq5} T=\frac{GM}{2\pi
R^2}. \ee Substituting (\ref{eq5}) into (\ref{eq3}), one obtains
the Newtonian force law as  the entropic force\be \label{eq6}
F=\frac{G m M}{R^2}. \ee

However, as was emphasized previously, the usage holographic screen
is not guaranteed to describe  a non-relativistic case of a source
mass $M$.

\section{Entropic force on the isothermal cavity }
In this section, to define a holographic screen,  we consider the
Schwarzschild spacetimes instead of the source mass $M$ as \be
\label{eq7} ds^2_{Sch}=g_{\mu\nu}dx^\mu
dx^\nu=-\Big(1-\frac{2GM}{r}\Big)dt^2+\frac{dr^2}{\Big(1-\frac{2GM}{r}\Big)}+r^2d\Omega^2_2.
\ee Here the event horizon (EH) is located at $r=r_{EH}=2GM$ whose
horizon area is  $A_{EH}=4\pi r_{EH}^2$.

It is well known that the Schwarzschild black hole could be  in
thermal equilibrium with a finite-size heat reservior in
asymptotically flat spacetimes. This is made by introducing a cavity
enclosed the black hole. In this case, we may introduce the local
temperature (Tolman temperature) and the quasilocal energy on the
isothermal cavity.  Now let us introduce the Tolman redshift
transformation on the black hole system~\cite{GPP2}. Using this
transformation, the local temperature observed by an observer
located at $r>r_{EH}$   is defined by~\cite{York} \be \label{eq8}
T_L(r)= \frac{T_{\infty}}{\sqrt{-g_{tt}}}=\frac{1}{8\pi G
M}\frac{1}{\sqrt{1-\frac{2GM}{r}}} \ee where \be \label{eq9}
T_\infty=\frac{1}{8\pi G M}=\frac{1}{4 \pi r_{EH}}\equiv T_H \ee is
the Hawking temperature $T_H$  measured by observer at infinity and
the denominator of $\sqrt{-g_{tt}}$ is the redshift factor. It is
worth to mention two limiting cases. On the cavity located  at
$r=r_{EH}+l_{pl}^2/r_{EH}$ near the event horizon, this local
temperature is given approximately by $T=1/4\pi l_{pl}$ which is
independent of the black hole mass $M$~\cite{Myungent}. On the other
hand, for $r \gg r_{EH}$, it reduces to the Hawking temperature
$T_H$.

Similarly, the quasilocal energy is derived  by considering the
first law of thermodynamics  and assuming that the
Bekenstein-Hawking entropy $S_{BH}=4\pi GM^2$ is not changed on the
cavity\be \label{loen} dE_{QL}(r)=T_{L}(r)dS_{BH}, \ee which is
integrated to give \be \label{quasile}
 E_{QL}(r)=\frac{r}{G}\Bigg[1-\sqrt{1-\frac{2GM}{r}}\Bigg].
\ee Here  the energy observed  at infinity is surely the ADM mass
$M$
 \be
 \label{loeni}
 E_{QL}(\infty)=M=\frac{r_{EH}}{2G}.\ee
Solving $E_{QL}$ for $M$ leads to \be
M=E_{QL}-\frac{GE_{QL}^2}{2r},\ee which states that the ADM mass
consists of the thermal energy and the gravitational self-energy to
create a cavity at $r>r_{EH}$. In this sense,  the isothermal
cavity, which was an artificial device to  make a phase transition
from a hot gas to a black hole~\cite{GPP2}, is different from the
holographic screen.

Consequently, it seems unlikely to define the entropic force on the
isothermal cavity except the  case that it located near the event
horizon.

 \section{Entropic force on the static holographic screen}

We briefly review how to  derive the entropic force on the static
holographic screen in the  Schwarzschild spacetime (\ref{eq7}). We
introduce the proper acceleration~\cite{Makela2} \be
a^\mu=\xi^\alpha\xi^\mu_{;\alpha}, \ee where $\xi^\alpha$ is a
timelike Killing vector field that satisfies
$\xi_{\mu;\nu}+\xi_{\nu;\mu}=0$. We define an integral \be
E(V)=\frac{1}{4\pi} \oint_{ \partial V} a^\mu n_\mu dA,\ee where $V$
is the bulk volume enclosed by spacelike hypersurface $\partial
V=S^2$ and $n_\mu$ is a spacelike unit normal vector $S^2$. This can
be rewritten as the Komar integral to define the concept of energy
in the stationary spacetime
 \be \label{KI} E(V)=\frac{1}{8\pi} \oint_{ \partial V}
\xi^{\mu;\nu} d\Sigma_{\mu\nu},\ee where \be d\Sigma_{\mu\nu}=\Big(
n_\mu \xi_\nu-n_\nu \xi_\mu\Big) dA. \ee For $r>r_{EH}$, a non-zero
component of the Killing vector is given by\be \xi^t=1. \ee On the
$S^2$, a non-zero component of the  normal vector is \be
n^r=-\sqrt{1-\frac{2GM}{r}}. \ee Using these, it is easily shown
that  the Komar integral (\ref{KI}) becomes  the local energy which
is defined on the holographic screen in the stationary
spacetime~\cite{Myungef} \be
E_{HS}=\frac{E_{\infty}}{\sqrt{1-\frac{2GM}{r}}},~~E_\infty=M.\ee
$E_{HS}$ indicates the energy of the gravitational field from the
viewpoint of an observer at rest with respect to the Schwarzschild
coordinate $r$.  It is important to note that the local energy
$E_{HS}$ defined on the holographic screen is not the quasilocal
energy $E_{QL}$ of (\ref{quasile}) on the isothermal cavity. At this
stage, we propose the local equipartition rule \be \label{eq11}
E_{HS}(r)=2S_{HS}(r)T_{HS}(r), \ee where the entropy $S_{HS}$ is
defined on the holographic screen located at $r
>r_{EH}$ \be \label{eq12} S_{HS}(r)= \frac{\pi r^2}{G}. \ee
 Then, the temperature on the holographic screen is given by
 \be
 \label{eq13}
 T_{HS}(r)=\frac{GM}{2\pi r^2 \sqrt{1-\frac{2GM}{r}}} \ee
 which takes a  different form from the local temperature $T_L(r)$ in
 Eq.(\ref{eq8}).
Plugging (\ref{eq13}) into (\ref{eq3}), we obtain the entropic
 force as
 \be
 \label{eq14}
 F_{HS}= 2\pi m T_{HS}(r)= \frac{1}{\sqrt{1-\frac{r_{EH}}{r}}} \frac{GmM}{ r^2} \ee
 which shows that  the mass $m$  feels an infinitely tidal force  in the limit of $r \to r_{EH}$,
 while it takes  the Newtonian force law at  the large distance of  $r\gg
 r_{EH}$. In this picture, it is hard to introduce a proper  acceleration
$a$ because one works in the stationary spacetime.

Consequently, in obtaining the entropic force (\ref{eq14}), an
important step was making use of the local equipartition rule
(\ref{eq11}) to assign the temperature on the static screen.

\section{Entropic force on the accelerating surface}

Now we are in a position to  define the entropic force on the
accelerating surface in the presence of the Schwarzschild black
hole (\ref{eq7}). In this case, the accelerating surface  plays
the role of the holographic screen. The accelerating surface was
introduced in \cite{Makela2,Makela3}. Here we use the accelerating
surface as the accelerating screen (AS).  For the non-stationary
spacetime, we introduce a future pointing unit vector \be u^\mu
\ee of the congruence for  the timelike world lines of the points
on $S^2$. It satisfies the orthogonality with the normal vector
$n_\mu$ as \be u^\mu n_\mu=0. \ee Also this vector is necessary to
define the change of the heat. In this case, a proper acceleration
vector field is defined by \be a^\mu=u^\alpha u_{;\alpha}^\mu \ee
and an AS is defined by the following properties: \be \sqrt{a^\mu
a_\mu}=a={\rm constant},~~a=a^\mu n_\mu\ee at every points on
$S^2$. The flux of the proper acceleration vector field through
the AS is defined to be \be \Phi_{AS}=aA_{AS},\ee where $A_{AS}$
is the area of the AS. A non-zero component of the  future
pointing unit vector
 is \be
u^t=\frac{1}{\sqrt{1-\frac{2GM}{r}}}, \ee and a non-zero component
of the proper acceleration vector $a^\mu$ is given by \be
a^r=u^tu^r_{;t}=-\frac{GM}{r^2}. \ee The proper acceleration defined
by  \be a=a^\mu n_\mu=
\frac{GM}{r^2}\frac{1}{\sqrt{1-\frac{2GM}{r}}} \ee plays the role of
the Unruh temperature \be T_U(r)=\frac{a}{2\pi}=\frac{GM}{2\pi
r^2}\frac{1}{\sqrt{1-\frac{2GM}{r}}}.\ee Considering the temperature
(\ref{eq13}) defined on the accelerating screen, we find that \be
T_{HS}=T_U\ee which means that the screen temperature $T_{HS}$ is
equal to the bulk temperature $T_U$.  A natural interpretation of
this result is that an accelerating observer on the AS observes
thermal radiation with the Unruh temperature $T_U$. This is a
well-known result of the local quantum field theory, which is called
the Unruh effect. This effect states that an accelerating observer
always observes thermal particles even when, from the viewpoint of
all inertial observers, there is no particles at all. In the
Newtonian limit, it reduces to \be a=\frac{GM}{r^2} \ee which is the
acceleration of particles in freely falling in the gravitational
field created by a point mass $M$. Then, the flux is given by \be
\Phi_{AS}=\frac{4G\pi M}{\sqrt{1-\frac{2GM}{r}}}, \ee when using
$A_{AS}=4\pi r^2$.

Now let us introduce the change of the heat  in terms of a
differential $d\Phi_{AS}$ of the flux $\Phi_{AS}$  through the AS as
\be \label{hf1} \delta Q= \frac{1}{4\pi G} d\Phi_{AS}. \ee
Considering that the surface gravity $\kappa$ is constant on the
event horizon, we keep $a$ fixed on the AS when varying the flux
$\Phi_{AS}$. That is, $da(r,M)=0$  implies \be dM=\frac{2GM
r-3(GM)^2}{r^2-GM r} \frac{dr}{G}. \ee Using this, Eq.(\ref{hf1})
leads to \be \label{hf2} \delta Q=\frac{2M}{r}
\frac{1}{\sqrt{1-\frac{2GM}{r}}}dr. \ee In this case, the entropy is
defined by \be S_{AS}=\frac{A}{2G}=\frac{2\pi r^2}{G}  \ee and thus,
its differential is $dS_{AS}=4\pi r dr/G$. Plugging this into
(\ref{hf2}) leads to the fact that the change of the heat is
balanced by the change of the entropy when fixing the temperature on
the AS \be
 \delta Q=\frac{GM}{2\pi r^2}
\frac{1}{\sqrt{1-\frac{2GM}{r}}}dS_{AS}=T_{U}(r)
dS_{AS}=\frac{a}{2\pi}dS_{AS}. \ee  We confirm that the entropic
force on the point mass $m$ near the AS is given by \be \label{hsf}
F_{HS}=2\pi m T_{U}=
ma=\frac{1}{\sqrt{1-\frac{2GM}{r}}}\frac{GmM}{r^2}. \ee Finally, we
observe that if one introduces the accelerating surface as the
accelerating screen, it is very natural to define the acceleration
$a$ of a test particle $m$ through the Unruh temperature $T_U$ and
thus, the entropic force (\ref{eq14}) is recovered properly.

 \section{Entropic force on the stretched  horizon}

 \begin{figure}[t!]
   \centering
   \includegraphics{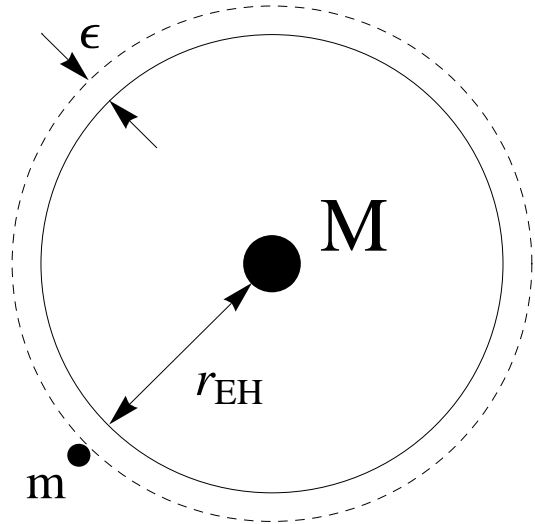}
\caption{The Schwarzschild black hole  is present by noting the
event horizon $r_{EH}$ (solid circle).  At the position
$r=r_{EH}+l^2_{pl}/r_{EH}$ near the event horizon, equality of
isothermal cavity=static holographic screen=accelerating screen is
depicted in terms of the dotted circle in order to define the
entropic force. } \label{fig.1}
\end{figure}

 Since the isothermal cavity, static holographic screen, and the
 accelerating screen do not provide a consistent form like the
 entropic force (\ref{eq6}), it is natural to ask where is the specific  place to give
 a consistent form.  In this section, we suggest that  this place is
just on the stretched
 horizon (SH)~\cite{SL}. This means that all thermodynamic
 quantities are measured by an observer located at the proper
 distance $l_{pl}$ away from the horizon.

 As is shown in Fig. 1, this horizon is considered as
the holographic  screen at $r=r_{EH}+l_{pl}^2/r_{EH}$ near the event
horizon. The length contraction of $
l_{pl}^2/r_{EH}=\sqrt{-g_{tt}}l_{pl}$  is due to the redshift
transformation of $\sqrt{-g_{tt}}\simeq l_{pl}/r_{EH}$ near the
horizon. On the SH, the local temperature is given by \be
T_L^{SH}=\frac{1}{4\pi
l_{pl}}\sqrt{1+\frac{l_{pl}^2}{r^2_{EH}}}\simeq \frac{1}{4\pi
l_{pl}}\Bigg(1+\frac{1}{2}\frac{l_{pl}^2}{r^2_{EH}}\Bigg)=\frac{1}{4\pi
l_{pl}}\Bigg(1+\frac{1}{8l_{pl}^2M^2}\Bigg)\ee which is independent
of the black hole mass $M$ in the leading order~\cite{Myungent}. On
the other hand, the HS temperature (=Unruh temperature) leads to \be
T_{HS}^{SH}=T_U^{SH}=\frac{1}{4\pi
l_{pl}\Big(1+\frac{l_{pl}^2}{r^2_{EH}}\Big)^{\frac{3}{2}}}\simeq
\frac{1}{4\pi
l_{pl}}\Bigg(1-\frac{3}{2}\frac{l_{pl}^2}{r^2_{EH}}\Bigg). \ee
Importantly, we observe that three temperatures are the same  \be
T^{SH}_{L}=T^{SH}_{HS}=T^{SH}_{U}=\frac{1}{4\pi l_{pl}} \ee in the
leading order.  Similarly, the local energy is given by \be
E_{HS}^{SH}=
\frac{r_{EH}^2}{2l^3_{pl}}\sqrt{1+\frac{l_{pl}^2}{r^2_{EH}}}\simeq
\frac{A_{EH}}{8\pi
l_{pl}^3}\Bigg(1+\frac{1}{2}\frac{l_{pl}^2}{r^2_{EH}}\Bigg). \ee In
the leading order, we check easily  that the equipartition rule is
satisfied as is shown by \be E_{HS}^{SH}= \frac{A_{EH}}{8\pi
l_{pl}^3}=\frac{N^{SH}T^{SH}_{HS}}{2},\ee with
$N^{SH}=A_{EH}/l_{pl}^2$. However, this shows a different case when
comparing with (\ref{eq4}) on the holographic screen for a
non-relativistic case.  Especially, the energy and temperature take
 different forms \be \label{let} E^{SH}_{HS}=2l_{pl}
M^2,~~T^{SH}_{HS}=\frac{1}{4\pi l_{pl}} \ee which show that the
energy is proportional to mass squared $M^2$, while the
temperature is independent of the mass.  We remind the reader that
this is a feature of thermodynamic quantities of black hole on the
stretched horizon, which can be approximated  by the Rindler
spacetimes where a commutation relation of $[M,t]=i$ is defined
for the Schwarzschild mass and time. Here, $E_{HS}^{SH}$ is the
Rindler energy  as defined by an observer near the horizon, while
$M$ is the Schwarzschild mass (energy) as measured by an observer
at infinity.  Importantly, we observe that the first law of
thermodynamics holds on the stretched horizon as \be
dE^{SH}_{HS}=T_{HS}^{SH}dS^{SH}. \ee Here the entropy on the
stretched horizon is given by \be S^{SH}=\frac{A_{EH}}{2l^2_{pl}},
\ee which is clearly  different from the the Bekenstein-Hawking
entropy \be S_{BH}=\frac{A_{EH}}{4l^2_{pl}}. \ee Finally, the
entropic force is consistently defined by \be \label{consif}
F_{SH}=2\pi mT^{SH}_L=2\pi m T^{SH}_{HS}=2 \pi m
T^{SH}_U=\frac{m}{2l_{pl}}, \ee which takes a familiar form of \be
F_{SH}=m a_{SH}\ee with the proper acceleration
$a_{SH}=\frac{1}{2l_{pl}}$ on the  stretched horizon.

However, we note that the quasilocal energy takes a different form
on the SH \be E^{SH}_{QL}\simeq \frac{1}{2l_{pl}
M}\Bigg(1-\frac{1}{2}\frac{l_{pl}^2}{r^2_{EH}}\Bigg) \ee when
comparing with the local energy in Eq.(\ref{let}). The equipartition
rule is not satisfied, as is evident from  $E^{SH}_{QL}\not= 2S_{BH}
T^{SH}_L$ in the leading order. Therefore, it seems that the
quasilocal energy is not directly  related to the entropic force.

\section{Entanglement system}

\begin{figure}[t!]
   \centering
   \includegraphics{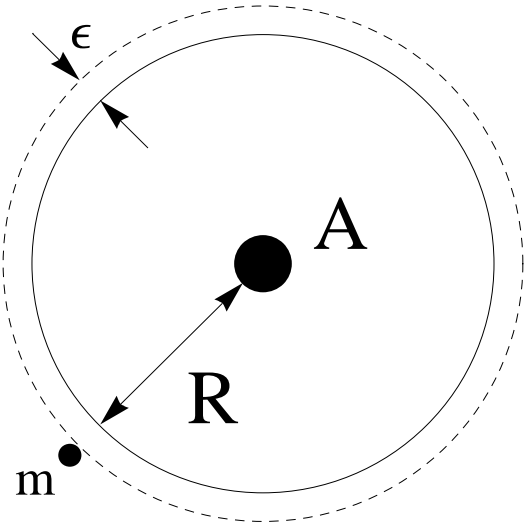}
\caption{The entanglement system  is present by noting the boundary
${\cal B}$ of area $A=4\pi R^2$ (solid circle) enclosed the volume
$V$. At the position $r=R+b^2/R$ near the boundary, a screen is
depicted in terms of the dotted circle to define the entropic force.
} \label{fig.2}
\end{figure}
Entanglement entropy is a general concept which is a coarse
graining entropy for a quantum system caused by an observer's
partial ignorance  of the information on state~\cite{BKLS}.  For
an entanglement system in the flat spacetime, we  consider the
three-dimensional
 spherical  volume $V$ and its enclosed boundary ${\cal B}=S^2$ (See Fig. 2).
 We assume that this system with
 radius $R$  and the cutoff scale $b$ is described by  the local quantum
 field theory of a free scalar field $\phi$.
The entanglement energy is carried by the modes around ${\cal B}$,
which implies that the cutoff scale $b$ is introduced only in the
$r$ direction through the length contraction $b^2/R^2$. We start by
noting the similarity between the entanglement system in the flat
spacetime and  the stretched horizon formulation of the
Schwarzschild black hole. Here we introduce the first law of
thermodynamics for the entanglement system \be \label{2eq1}
dE_{ENT}=T_{ENT}dS_{ENT},\ee whose thermodynamic quantities are
given by~\cite{MSK1}\be \label{2eq2}
 S_{ENT} \simeq \frac{A}{b^2},~ E_{ENT}\simeq \frac{A}{ b^3},~ T_{ENT}\simeq \frac{1}{ b},  \ee
 where $A=4\pi R^2$ is the
 area of the  boundary $ {\cal  B} $ of the system.
Eq.(\ref{2eq2}) shows a universal behavior for all entanglement
systems in Minkowski spacetime. We note that the zero-point energy
of the system was subtracted in the calculation of the entanglement
energy $E_{ENT}$, thus  degrees of freedom  on the boundary
contribute to giving  $E_{ENT} \sim A$. It seems that its areal
behavior  is compatible with the concept of the entanglement.
 The entanglement entropy
behaves universally as $S_{ENT} \sim A$, which takes the same form
as that of  the black hole. This is why the entanglement entropy is
considered as the origin of the black hole entropy. The entanglement
temperature is independent of the radius $R$ of system. Considering
the connection between $R \leftrightarrow r_{EH}$ and $b
\leftrightarrow l_{pl}$, thermodynamic quantities   are the nearly
same as those of the black hole on the stretched horizon: \be
S_{ENT} \leftrightarrow S^{SH},~~E_{ENT} \leftrightarrow
E^{SH}_{HS},~~T_{ENT} \leftrightarrow T^{SH}_{HS}=T^{SH}_L=T^{SH}_U.
\ee

In order to compare (\ref{2eq2})  with those for the black hole
observed at infinity,  one observes that  entanglement quantities
at $r=R+b^2/R$   are blueshifted with respect to those at infinity
by inserting the factor of $1/\sqrt{-g_{tt}}$.  With
$\sqrt{-g_{tt}}\simeq b/R$, one obtains $E_{ENT}^{\infty}\simeq
\sqrt{-g_{tt}}E_{ENT}=R/b^2$ and $T_{ENT}^{\infty}\simeq
\sqrt{-g_{tt}}T_{ENT}=1/ R$~\cite{Myungent,MSeo}. It shows how an
inclusion of gravity alters thermodynamics of the entanglement
system.   Later on, the same authors have calculated the
entanglement energy in the Schwarzschild background without
introducing   the redshift factor~\cite{MSK2}.

Comparing the entanglement system with the stretched horizon
 of the Schwarzschild black hole, two are very similar to
each other even though the entanglement system does not have the
source  mass $M$ and thus, it has no  effects of  gravity
definitely.   Importantly, we may consider the entanglement system
as a genuine system to define the entropic force. This is because we
could define the entropic force in the entanglement system without
introducing the mass $M$.  In this case, the equipartition rule \be
E_{ENT}\simeq  N_{ENT}T_{ENT} =\frac{A}{b^2}\frac{1}{ b} \ee is very
special because the energy is proportional to the area $A$ (but not
the mass $M$) and the temperature is independent of the area.

Finally, we propose that an entropic force could be defined by \be
\label{entfo} F_{ENT}=2\pi m T_{ENT}\simeq \frac{m}{2b}, \ee which
takes another form of \be F_{ENT}\simeq m a_{ENT}\ee with the proper
acceleration $a_{ENT}=\frac{1}{2b}$ in the flat spacetime.  This
proposition may be a good idea  to ensure the entropic force if the
latter does really exist.

\section{Discussions}

It is fare to say that the origin of the gravity is  not yet fully
understood. If the gravity is not a fundamental force, it may be
emergent from the other approach to gravity.  Newtonian force law
could be emergent from the equipartition rule and the holographic
principle~\cite{Ver}. However, one important issue is whether the
holographic screen could be nicely defined for a non-relativistic
case of a source mass $M$.

In this work, in order to realize  the entropic force, we have first
introduced the isothermal cavity, static holographic screen, and
accelerating screen as a candidate for the holographic screen by
implementing the Schwarzschild spacetimes. Then,  using
Eq.(\ref{eq3}), we have found  the entropic force (\ref{eq14}) and
(\ref{hsf}) which are different from the original form (\ref{eq6}).
Furthermore, these might merge to provide a unified picture to
define the entropic force (\ref{consif}) on the stretched horizon
near the event horizon.

It is well-known that the entanglement system also respects the
equipartition rule and the holographic principle in the flat
spacetime~\cite{MSK1}.
 Considering a close relationship  between black hole
thermodynamics on the stretched horizon and the entanglement
system, we have proposed that an entropic force could be realized
as (\ref{entfo})  in the entanglement system   without referring a
source mass $M$.

Finally, we should say that our approach is not a complete scheme to
understand the entropic force because we have already introduced a
given configuration of gravitational field (Schwarzschild
spacetimes),  to define the holographic screen. However, given the
holographic screen properly as was in this work, one could define
the entropic force by using (\ref{eq3}).

\section*{Acknowledgment}
The author  thanks Rong-Gen Cai for helpful discussions on an
entropic force. This work  was supported by Basic Science Research
Program through the National  Research  Foundation (NRF) of  Korea
funded by the Ministry of Education, Science and Technology
(2009-0086861).

\end{document}